\documentclass[11pt]{article}
\usepackage{moriond,epsfig,natbib,journals}
\def\be{\begin{equation}}
\def\ee{\end{equation}}
\def\bea{\begin{eqnarray}}
\def\eea{\end{eqnarray}}

\begin{document}
\vspace*{4cm}
\title{A COMPARISON OF THE EUV AND X-RAY EMISSION PROFILES IN SIX
CLUSTERS OF GALAXIES}

\author{ Florence DURRET}
\address{Institut d'Astrophysique de Paris, 98bis Bd Arago, 75014 Paris,
France}

\author{ Sergio DOS SANTOS}
\address{Department of Physics and Astronomy, Leicester University, University
Road, Leicester LE1 7RH, UK}

\author{ Richard LIEU}
\address{Department of Physics, University of Alabama, Huntsville AL 35899, 
USA}

\maketitle\abstracts{
We have performed the wavelet analysis and image reconstructions 
for six clusters of galaxies observed both at extreme ultraviolet and
X-ray energies. The comparison of the profiles of the EUV and X-ray
reconstructed images shows that they differ at a very large confidence
level in five of the six clusters, and are similar only for the Coma
cluster, which is by far the hottest one in the sample. }

\section{Introduction}

The Extreme Ultraviolet Explorer (EUVE) has detected emission from a
few clusters of galaxies in the 70-200 eV energy range. By order of
discovery, these are: Coma \citep{Lieu1}, Virgo
\citep{Lieu2,Berghofer1}, A~1795 \citep{Mittaz}, A~2199
\citep{Bowyer}, A~4059 \citep{Berghofer2} and Fornax (Bergh\"ofer,
Bowyer \& Korpela 2001). The problem of the physical origin of the EUV
emission was immediately raised. Excess EUV emission relative to the
extrapolation of the X-ray emission to the EUVE energy range was
detected in several clusters, suggesting that thermal bremsstrahlung
from the hot (several $10^7$ to $10^8$K) gas emitting in X-rays could
not be entirely responsible for the EUV emission. However, the reality
and importance of this effect is still controversial.

This excess UV emission has been interpreted as due to two different
mechanisms: thermal radiation from a warm ($10^5-10^6$K) gas, as first
suggested by Lieu et al. (1996a), or inverse Compton emission of
relativistic electrons either on the cosmic microwave background or on
stellar light originating in elliptical galaxies
\citep{Bowyer1,Sarazin}.  Although the existence of a multiphase
intracluster medium is likely \citep{Bonamente1,Lieu4}, there are
several difficulties with the thermal model. The most serious one is
that since gas in the temperature range $10^5-10^6$K cools very
rapidly, a source of heating (and refuelling) for this gas is
necessary, and no satisfactory model has been proposed until now to
account for the gas heating.  On the other hand, non thermal models
\citep{Sarazin, Bowyer1,Lieu3,Ensslin,Atoyan,Brunetti} seem able to
account for the excess EUV emission, but still require some fine
tuning.

We have performed for a sample of six clusters a wavelet analysis and
reconstruction of the EUVE and ROSAT (PSPC for five of them and ROSAT
HRI for A~4038) images.  We compare the profiles of the images
thus obtained, arguing that if both emissions have the same physical
origin the profiles should be identical.

\section{The data and data analysis}

\subsection{The data}

Observations in the EUV range were made with the EUVE satellite.  The
ROSAT data were retrieved from the archive.  The exposure times and
main cluster characteristics are given in Table~\ref{tbl:mainchar}.

\begin{table}[h]
\centering
\caption{Exposure times and main cluster characteristics.}
\begin{tabular}{|lrrclrrc|}
\hline
Cluster & EUVE total  & ROSAT  exp. & ROSAT & Redshift & kT$_{\rm X}$ & Scaling & Cooling\\
         & exp. time (s)  &  time (s)   &  ObsId       &          & (keV)        & (kpc/superpx) & flow\\
 \hline		    
 A 1795  & 158689    &   33921     & rp800105     & 0.063    & 5.9          & 32.7    & yes \\
 A 2199  &  93721    &   34633     & rp800644     & 0.0299   & 4.1          & 15.8    & yes \\
 A 4038  &  161492   &   11480     & rh800962     & 0.0283   & 4.0          & 14.9    & yes \\
 A 4059  &  145389   &    5225     & rp800175     & 0.0478   & 4.0          & 25.0    & yes\\
 Virgo   &  146204   &    9135     & rp800187     & 0.003    & 2.4          &  1.6    & weak \\
 Coma    &   60822   &   20112     & rp800005     & 0.023    & 8.7          & 12.2    & no \\
\hline
\end{tabular}
\label{tbl:mainchar}
\end{table}

In order to have comparable spatial resolutions in the EUV and X-rays,
the EUV images were rebinned 4$\times$4, leading to a ``superpixel''
size of 0.307 arcminutes (18.4 arcseconds). Reduced X-ray images were
also rebinned (from 15 arcsec pixels for the PSPC and 5 arcsec pixels
for the HRI) to the same 18.4 arcsec pixel size in order to allow a
direct comparison of the EUV and X-ray profiles.

The linear scale at the cluster distance per superpixel, estimated
with H$_0=50$~km~s$^{-1}$~Mpc$^{-1}$ and q$_0$=0 is given in column 7
of Table~\ref{tbl:mainchar}. 

\subsection{Data reduction}

The {\sf ROSAT\/} X-ray data for each cluster were obtained from the
HEASARC web archive\footnote{http://heasarc.gsfc.nasa.gov/}. Because
of its higher sensitivity, we priviledged, when possible, PSPC data
over HRI and pointed observations over all-sky survey scanning mode
observation. When multiple PSPC observations were available, we
always used the one with the smallest offset radius compared to the
center of the cluster, and the longest exposure.

The ROSAT PSPC data were reduced with S. Snowden's Extended source
Analysis Software.  We limited our analysis to the [0.5,2.0] keV
energy band \citep[bands R4-R7 as defined in][]{Snowden}. This
preliminary reduction produced a surface brightness image with $512
\times 512$ pixels, 15 arcsec per pixel (roughly the FWHM of the
{\sf PSPC} PSF at $1 {\rm keV}$), as well as an exposure image for
each of the bands and a background image.

The ROSAT HRI data for A4038 were reduced with the same ESAS package,
leading to an image of $512 \times 512$ pixels with 5 arcsec per
pixel.

The EUVE/Deep Survey photometer (DS) images where extracted from the
raw event lists with the customary linear bin size of 13 pix/arcmin.
The rectangular shape of the Lex/B filter ($\sim$70-200 eV) results in
highly elongated images (see Fig. 2).  When more than one observation
was available for each target, images where coadded to improve the
S/N; the rectangular shape often results in only partial overlap
between the exposures (e.g. Fig. 3).

\subsection{The wavelet transform}

The detailed method and its tests on simulated and real images will be
presented elsewhere (Dos Santos et al., in preparation). Here, we will
only highlight the different steps.

The wavelet transform (hereafter WT) has proven its capabilities in
numerous astronomical applications, such as the detection of the large
scale structure ({\it e.g.} Slezak 1993), galaxy detections and counts
\citep{Slezak90}, reconstruction of cluster X-ray images
\citep{Slezak94} and structure detection in low intensity X-ray images
({\it e.g.} Dos Santos \& Mamon 1999). We used the {\it \`a trous}
discrete wavelet algorithm \citep{Shensa92}, implemented in a package
kindly provided by E. Slezak.  In the {\it \`a trous} implementation
of the WT, an $N \times N$ image is transformed into $P$ wavelet
planes (hereafter WP), each being the difference between two
consecutive wavelet convolutions at scales $i$ and $i+1$ (with $2^i$
and $2^{i+1}$ pixels respectively, with $1 \leq i \leq P$). The pixel
values in the plane $i$ are called the wavelet coefficients at scale
$i$. The main advantage of this algorithm is that each WP has the same
number of pixels as the initial image, and thus, the reconstruction of
the image (for example after thresholding in the WPs) is a
straightforward addition process.

The noise estimation in the WPs is the main difficulty of the
reconstruction process. Once the noise in the WPs has been estimated,
it is straightforward to identify the wavelet coefficients which are
statistically significant at a certain level. These coefficients are
left unchanged, while all the others are replaced by zero (this
process is often called ``hard thresholding''). Then a simple addition
of the thresholded WPs gives what is called the denoised image, which 
will be used for scientific analysis.

\subsection{Estimation of the noise in the Wavelet Planes}
\label{sec:estnoise}

For an image with a stationary gaussian white noise, the resulting noise in
each WP is itself gaussian, and its standard deviation can be computed
analytically, once the resultant of a gaussian noise of standard deviation 1
has been computed on each WP \citep[see {\it e.g.\/}][]{Rue}. 

For Poisson noise, techniques of variance stabilization have been
introduced, which work under the assumption that the mean value of the
image is large \citep[typically, more than 30 counts per pixel, see
][]{Murtagh95}. However, for very low counts (typically less than 10),
this Anscombe transform \citep{Anscombe} looses control over the
bias. In this case, an alternative approach is needed. 

We have adopted a new method for the noise evaluation in each WP. The
complete method will be presented elsewhere (Dos Santos et al., in
preparation), and we only outline the important features here.

Basically, what we want to assess is the probability that a certain wavelet
coefficient $c_i(x,y)$ (at scale $i$ and position $(x,y)$) is due to
noise. We first consider an initial image which is zero-valued everywhere,
except on its central pixel, where its value is $p(x_0,y_0)$ (the translation
invariance property of the wavelet transform ensures that choosing the
central pixel has no consequences on the generality of the reasoning). After
the wavelet transform of the image, this single pixel will have a
non-vanishing contribution on each of the WPs. Of course, since the wavelet
function has a finite support, the contribution of this single pixel outside
this support will be zero. Suppose now that we have an estimation of the
noise level on this particular pixel. This estimation is $n(x_0,y_0)$. We can
then simulate an image with, say, Poisson noise (or whatever noise process we
choose) of mean $n$ and apply to it the wavelet transform. Looking at the
histogram of each WP of the simulated image in turn, we can measure the
minimum threshold $n_i$ that ensures that a fraction F of the wavelet
coefficient pixels of plane $i$ have a value lower than $n_i$. In other
words, a value of a pixel in the WP $i$ will have a probability F to be {\it
lower} than $n_i$ (or a probability 1-F to have a value {\it higher} than
$n_i$). Now returning to the initial image and its wavelet transform, we can
compare the value of each wavelet coefficient in plane $i$ with this
threshold $n_i$. If this value is higher, its probability of not being
due to noise is 1-F and we keep it. If it is lower, we don't keep it and put
its value to zero.  We do this in turn for each WP. Since the pixels are
considered as independent, we can repeat the same simulation and the same
thresholding on each pixel of an image (which has got many non-zero valued
pixels this time). Having an estimation of the noise for each pixel, we
construct, for each of these pixels, a simulated image having a mean value
equal to this level of noise, and compute its wavelet transform, the
thresholds for each wavelet plane, and then replace by zero the wavelet
coefficients which are lower than this threshold.

In order to estimate the noise level on each pixel, we will suppose
that each pixel is in fact {\it entirely due to noise}. The obvious
estimation of the noise level is then the value of the pixel
itself. All we have to do then is, for each pixel value in the initial
image, to simulate an image with noise having as a mean the pixel
value itself. After computing the thresholds in the WPs of the
simulated image, we compare them to the value of the wavelet
coefficient in each WP of the initial image at the position of the
pixel and see if the real values are lower than the threshold.

In fact, this method is conservative in the sense that it tests the
hypothesis that {\it all the pixels in the real image are due to
noise}. The value of a wavelet coefficient in a given WP at a certain
position will result from the addition of the wavelet transform of all
the pixels of the image. Locally, one assumes that the image is in
fact a flat noisy image having a level equal to the pixel value ({\it
that is what we assume in the simulated images}). If, surrounding a
pixel, other pixels are not due to noise, their wavelet coefficient
contribution at the position of the pixel will exceed the noise
contribution, and the wavelet coefficient at the position of the pixel
will be higher than the threshold measured in the simulations.

\begin{figure}
\begin{center}
\psfig{figure=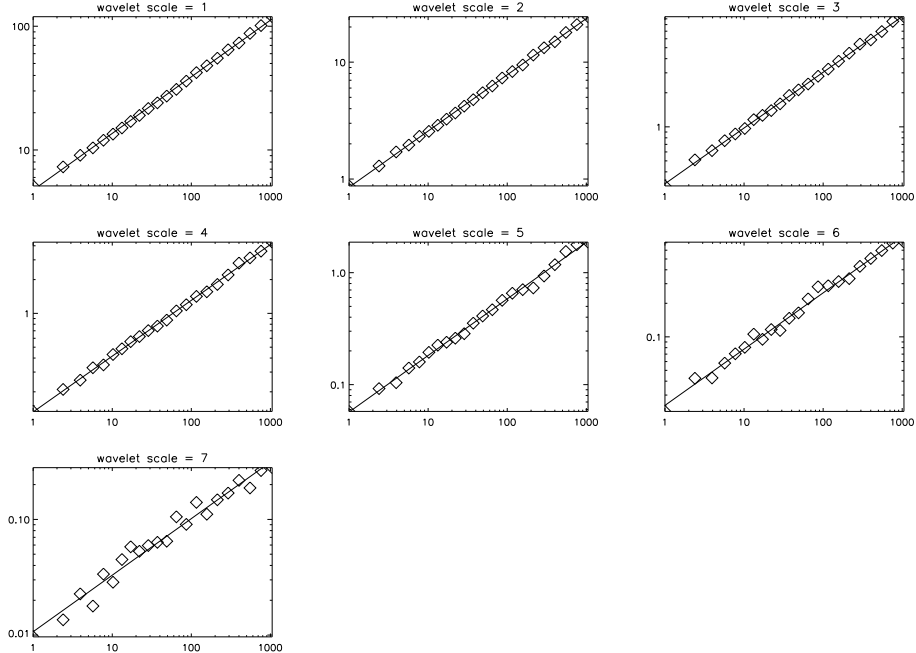,height=9truecm}
\caption{Fit of the WP thresholds as a function of the pixel 
value.
\label{fig:wp}}
\end{center}
\end{figure}

In practice, we don't take as the local noise estimation the value of
the pixel itself, but the median value over a $3 \times 3$ pixel square
surrounding the pixel. This ensures that we don't have spurious
zero-valued pixels surrounded by non-zero pixels. Second, we don't do
a simulation for each pixel value of the image, but take the minimum
and maximum values in the image, and choose N levels logarithmically
between these two values (typically, $N \sim 10$). We simulate N
images with these N mean values and plot the threshold against the
pixel value for each wavelet plane. We then fit the relation
threshold-pixel value in each WP and use this fitted law to estimate
the threshold in each WP for each pixel of the image. 
Fig.~\ref{fig:wp} shows this relation in logarithmic coordinates, as
well as the fit.

\subsection{Final image reconstruction}

The final image reconstruction is a simple process of addition of the
thresholded WPs. We don't iterate for this reconstruction, as this
proves to be very time-consuming, without adding much information. We
will show elsewhere (Dos Santos et al., in preparation) that this does
not introduce important artefacts in the reconstructed image, although
the hard-thresholding in the wavelet planes can sometimes cause
changes of slopes in the reconstructed images, on the boundary from
one scale to another.

\subsection{Avoiding edge effects in the EUVE images}

EUVE images exhibit the particular rectangular geometry of the
detector. The fact that the length of one side greatly exceeds that of
the other has proven to be a problem for the wavelet reconstruction
algorithms, because the number density of counts goes rapidly to zero
at the edge of the detector in one direction, but not in the other
direction. In the wavelet planes, this large gradient translates in
large wavelet coefficient values, particularly on the smallest
scales. This in turn produces false object detections on a scale
comparable to the photon count decrease.

The first simple idea to deal with this problem is to cut the image in
order to have a comparable length and width, but the cluster emission
is obviously truncated by this method, and the emission of all the
clusters then doesn't extend further than 5 arcmin.  This is obviously
much smaller than detected by any classical analysis ({\it i.e.\/,}
counts in radial bins), even the most conservative.  We thus had to
design a more complicated algorithm, in order to avoid false object
detection without artificially limiting the extent of real cluster
emission.

Fig.~\ref{fig:virgo} shows the rebinned EUV image of the Virgo
cluster. The cluster is clearly visible in the middle of the field of
view.  Three rotated rectangles are superimposd on the raw image. The
largest one (ABCD) follows the outer detector shape closely; its edges
are detached from the detector boundaries to avoid the steepening of
the background near the edge, which would bias the noise estimation
\citep[see Fig.~2 of][]{Lieu6}. In this box, we define two
rectangles (denoted $r_1$ and $r_2$), one on each side, as far away
from the central source as possible. Both $r_1$ and $r_2$ are assumed
to contain only noise and to be representative of the noise level. The
second order moment of the count distribution is used to estimate the
noise level in each of these rectangles. The noise level in the large
box is then computed as the mean of the noise levels in both
rectangles. Each pixel outside the larger box is then replaced by a
random number drawn from a Poisson distribution, the variance of which
is obtained by adding in quadrature the variances in each
rectangle. The accuracy of the noise estimation is readily observable
in the product image : if the noise is {\it underestimated}, the
central box will obviously rise against the rest of the image, with
its boundaries showing a clear discontinuity. On the other hand, if
the noise has been {\it overestimated}, the outside parts of the new
image rise against the central box, again showing a discontinuity on
the boundaries. This particular technique allows us to recover a flat
asymptotic background far from the center, a necessary condition for
the good reconstruction and denoising of the image
\citep{Slezak94}. The wavelet-based reconstruction algorithm presented
in the last section is applied to this newly created image.

\begin{figure}
\begin{center}
\caption[Virgo polygons]{Virgo long exposure -- definition  of boxes in
order to obtain the image ready for wavelet reconstruction. Note that
the different boxes here are only illustrative of the algorithm we used. }
\label{fig:virgo}
\end{center}
\end{figure}

When more than one EUVE observation of a cluster was available, we
used the co-added image in order to increase the signal-to-noise
ratio. Fig.~\ref{fig:a1795box} illustrates the case of A1795. This
image is typical of the co-added images we dealt with in the case of
other clusters. The difference in roll-angle of the satellite in both
observations is apparent. The central intersection of both
observations obviously has the highest signal-to-noise ratio. For each
of the separate observations, we draw again a box that follows
the outer detector shape closely. These boxes are labelled ABCD and
A$^\prime$B$^\prime$C$^\prime$D$^\prime$ in
Figure~\ref{fig:a1795box}. In each of the boxes, we define two
rectangles, where the noise is evaluated. The noise level in a box is
computed as the mean of the noise levels in both rectangles. Once this
has been done for each box, each pixel outside the intersection of the
boxes (the central diamond) is replaced by a random number drawn from
a Poisson distribution, the variance of which is obtained by adding in
quadrature the variances in each box. Once again, the under or
overestimation of the noise will be clearly visible, and we apply to
this image the reconstruction algorithm.

\begin{figure}
\begin{center}
\caption[A1795 polygons]{A 1795 definition of boxes in order to obtain
the image prior the wavelet reconstruction in the case of a co-added
image. } 
\label{fig:a1795box}
\end{center}
\end{figure}

Obviously, we won't detect any structure outside a certain radius
since the outer part of the image is made of pure noise. The central
circle in Fig.~\ref{fig:a1795box} shows a conservative estimate of
this radius ($\sim$17 arcminutes). We will thus compare X-ray and EUVE
brightness profiles only inside this radius.

Dealing with ROSAT images does not present the same problems. The only
geometrical complication arises from the presence at a radius of
approximately 20 arcmin of the structure support of the
telescope. This is not a big issue since the limiting radius obtained
from the EUVE images is usually smaller than this value. When this is
not the case ({\it e.g.} for Coma), this ring is taken as the limiting
radius. On the other hand, what is of great concern here is the noise
probability distribution function (hereafter PDF) after the data
reduction. Indeed, the noise PDF is poissonian only in the raw counts
detected by the PSPC, before Snowden's algorithms are applied. The
exposure correction, as well as the subtraction of the different
modelled backgrounds change drastically the noise PDF, which cannot be
approximated any more by a Poisson process. Consequently, we have to
take this fact into account in the reconstruction algorithm.

The images obtained after data processing are exposure-corrected and
background-subtracted.  We obtain the estimate of the various
backgrounds (particle background, scattered solar X-ray background,
...) in sky coordinates, {\it i.e.\/,} in images with dimensions equal
to those of the corrected image. An exposure image is also produced by
the ESAS package. Adding the background files to the final image, we
obtain a non-background-subtracted image. We need to run the ESAS
package in order to have the background estimates, as well as the
exposure maps in the different bands (which are summed after weighting
by the total number of counts in each band). It is from this image
that we compute our simulated images. Namely, as in the traditional
reconstruction algorithm, we choose $N$ levels between the minimum and
the maximum pixel value of the image, and construct $N +1$ uniform
images with a number of counts per pixel equal to each of these
levels. Each of these images is then multiplicated by the mean
exposure map, and then only the Poisson noise is added to the
simulated image (the noise is thus added to the non-exposure-corrected
image, {\it thus simulating the observation before the data
reduction).} After this, we divide the noisy simulated image by the
mean exposure map and compute its wavelet transform. The rest of the
algorithm is analogous to that presented in
section~\ref{sec:estnoise}. This supplementary step in the
construction of the simulated images allows us to simulate optimally
the process of observation and data reduction of the ROSAT files.

\subsection{Radial profiles}

Once both (EUVE and ROSAT) images of a cluster were
wavelet-reconstructed, we derived profiles from these 2D reconstructed
images within elliptical concentric rings using the
STSDAS.ANALYSIS. ISOPHOTE.ELLIPSE task in IRAF.  For each cluster, the
ellipticity and major axis position angle were first determined for
the X-ray images, then fixed for both images, in order to compare the
X-ray and EUV fluxes in the same spatial regions.  Error bars on the
profiles were computed by calculating the number of counts in each
ring and taking as the error the square root of this number of counts.

\section{Results}

\begin{table}
\centering
\caption{Radius of the EUV and X-ray emissions derived from the wavelet 
reconstructed images. If the image is not circular, two numbers are given:
the first is the horizontal (East-West) radius, the second the vertical 
(North-South) radius.}
\begin{tabular}{|lrrrr|}
\hline
Cluster & EUV Radius & X-ray radius & EUV Radius & X-ray radius \\
        & (arcmin)   & (arcmin)     & (kpc)      & (kpc) \\
\hline		    
A 1795  & 17       & 16      & 1800     & 1700  \\
A 2199  & $\geq$17 & 18      & $\geq$880 &  940 \\
A 4038  & 4.0-6.5  & 7.8-6.5 & 190-310  & 380-340   \\
A 4059  & 7.4-5.7  & 17      & 600-460  & 1350  \\
Virgo   & $\geq$17 & 18      & $\geq$90 & 100 \\
Coma    & $\geq$17 & 20      & $\geq$660 & 780 \\
\hline
\end{tabular}
\label{tbl:size}
\end{table}

\begin{table}
\centering
\caption{Kolmogorov Smirnov test on EUV and X-ray profiles: probability 
P that the profiles differ.}
\begin{tabular}{|lc|}
\hline
Cluster & P \\
\hline		    
A 1795  & 0.99940 \\
A 2199  & ($1-9\ 10^{-10}$) \\
A 4038  & 0.9990 \\
A 4059  & ($1-3\ 10^{-8}$) \\
Virgo   & ($1-3\ 10^{-6}$) \\
Coma    & 0.1916 \\
\hline
\end{tabular}
\label{tbl:ks}
\end{table}

The maximum extents detected for the EUV and X-ray emissions are given
in Table \ref{tbl:size} for each cluster.  The results of a
Kolmogorov-Smirnov (hereafter K-S) test estimating for each cluster
the probability that the EUV and X-ray profiles differ are given in
Table \ref{tbl:ks}.

The EUV and X-ray profiles for the six clusters are displayed in
Figs. \ref{fig:a1795profil} to \ref{fig:comaprofil}. The
profile comparison for each cluster is briefly presented below.

The EUV and X-ray radii of A~1795 are 17 and 16 arcmin respectively
(Fig. \ref{fig:a1795profil}), corresponding to physical extents of
1800 and 1700 kpc. These extents are notably larger than previously
found \citep{Mittaz,Bonamente1} and are by far the largest in all the
sample (in physical units). The EUV and X-ray profiles appear to
differ quite markedly, as confirmed by the K-S test, which gives a
probability of 0.9994 that the two distributions are different (see
Table \ref{tbl:ks}).

\begin{figure}
\begin{center}
\psfig{figure=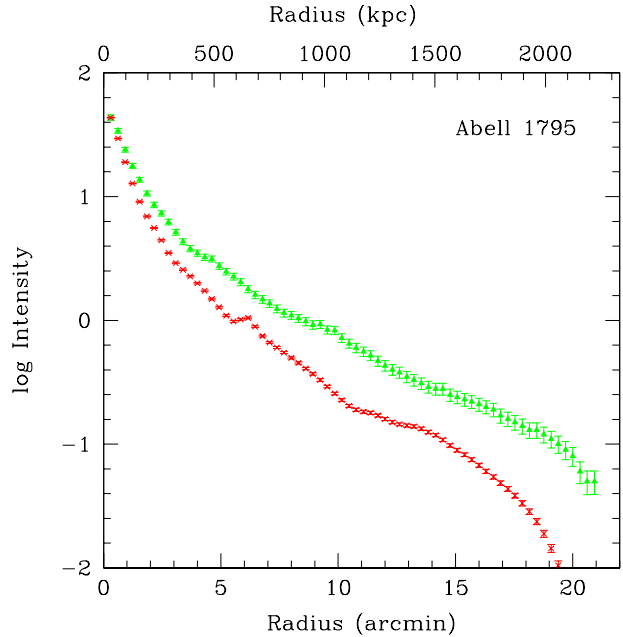,height=9truecm}
\caption[a1795_Ir_EUVX_fix_log.ps]{EUV (green triangles) and X-ray 
(red crosses) profiles for A~1795, with the radius expressed in 
arcminutes (bottom) and kpc (top) as in the five following figures. }
\label{fig:a1795profil}
\end{center}
\end{figure}

The EUV and X-ray radii of A~2199 are 20 and $\geq$17 arcmin (our
conservative limit, as explained in section 2.6) respectively
(Fig. \ref{fig:a2199profil}), or $\geq$880 and 940 kpc, larger than
measured by \citet{Kaastra} both with EUVE and Beppo-Sax. Here also,
the profiles differ with a very high probability.

\begin{figure}
\begin{center}
\psfig{figure=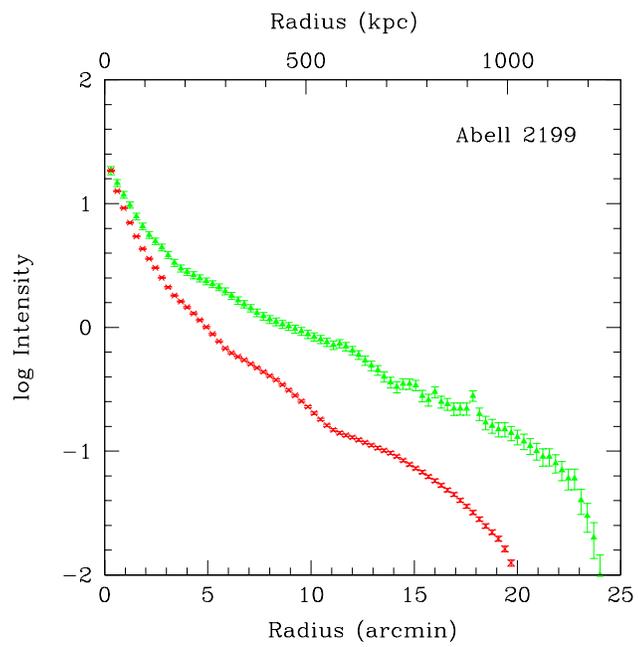,height=9truecm}
\caption[a2199_Ir_EUVX_fix_log.ps]{EUV and X-ray profiles
for A~2199. }
\label{fig:a2199profil}
\end{center}
\end{figure}

A~4038 is located very close to a second EUV source, making it
difficult to draw a profile for the EUV emission. We crudely modeled
the EUV emission of this second source and subtracted it from the
image, then deriving a profile for A~4038. This cluster is found to
have the smallest radial extent in our sample, both in EUV and X-rays:
4.0-6.5 and 7.8-6.5 arcmin respectively (see Table \ref{tbl:size} and
Fig. \ref{fig:a4038profil}). These values correspond to 190-310 and
380-340 kpc. This is the only cluster in our sample for which we used
HRI data; due to the lower sensitivity of the HRI, it is not
surprising that its X-ray emission is not detected very far from the
center (less than 400 kpc in physical units). The EUV and X-ray
profiles differ with a high probability.

\begin{figure}
\begin{center}
\psfig{figure=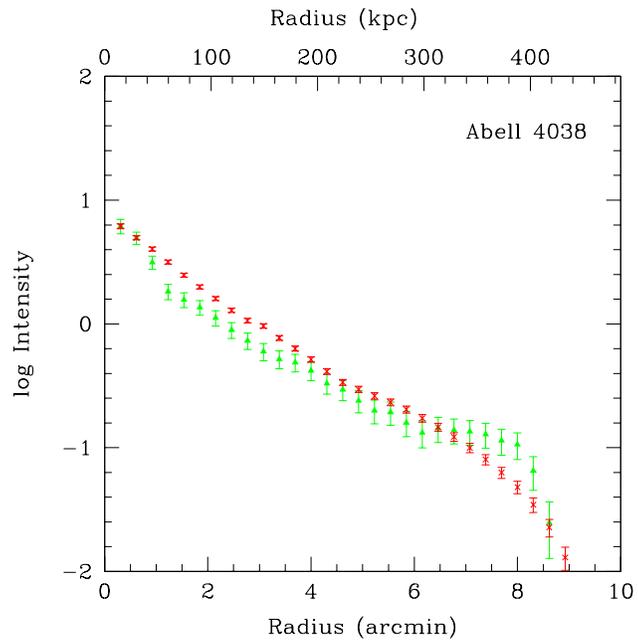,height=9truecm}
\caption[a4038_Ir_EUVX_fix_log.ps]{EUV and X-ray profiles
for A~4038. }
\label{fig:a4038profil}
\end{center}
\end{figure}

In A~4059, the EUV and X-ray emissions reach 7.4-5.7 and 17 arcmin
respectively (Fig. \ref{fig:a4059profil}), somewhat larger than
reported by \citet{Berghofer1} and corresponding to 600-460 and 1350
kpc. The EUV and X-ray profiles differ with a very high probability.

\begin{figure}
\begin{center}
\psfig{figure=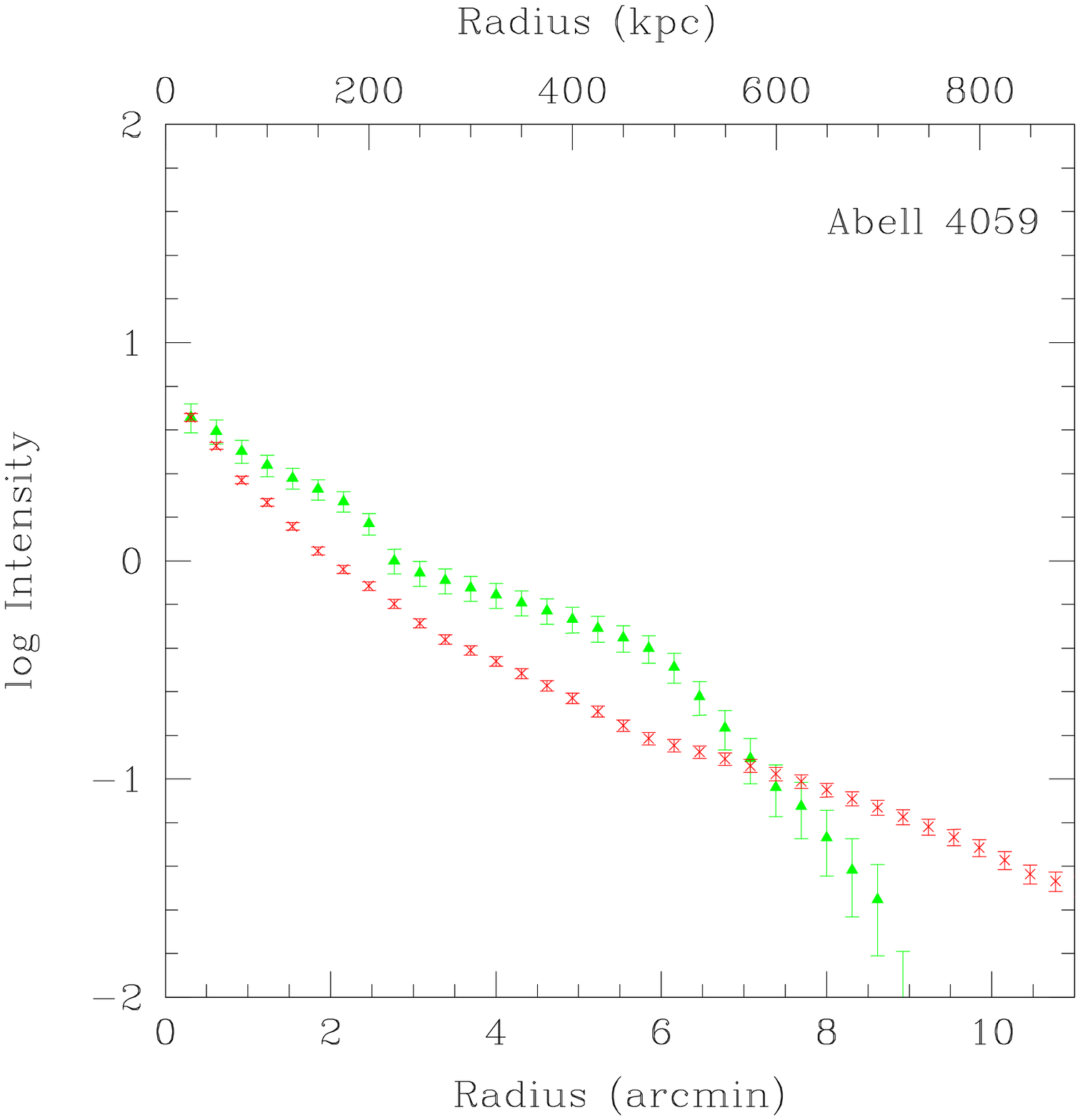,height=9truecm}
\caption[a4059_Ir_EUVX_ell_logn2.ps]{EUV and X-ray profiles
for A~4059. }
\label{fig:a4059profil}
\end{center}
\end{figure}

Virgo is the coldest and most nearby cluster in our sample. EUV and
X-ray emissions reach $\geq$17 and 18 arcmin respectively
(Fig. \ref{fig:virgoprofil}), that is larger than previously detected
\citep{Lieu2,Berghofer2,Bonamente1}; however, these extents correspond
to the very small values of 90 and 100 kpc, due to the small redshift
of Virgo. Here also, the EUV and X-ray profiles differ with a high
probability.

\begin{figure}
\begin{center}
\psfig{figure=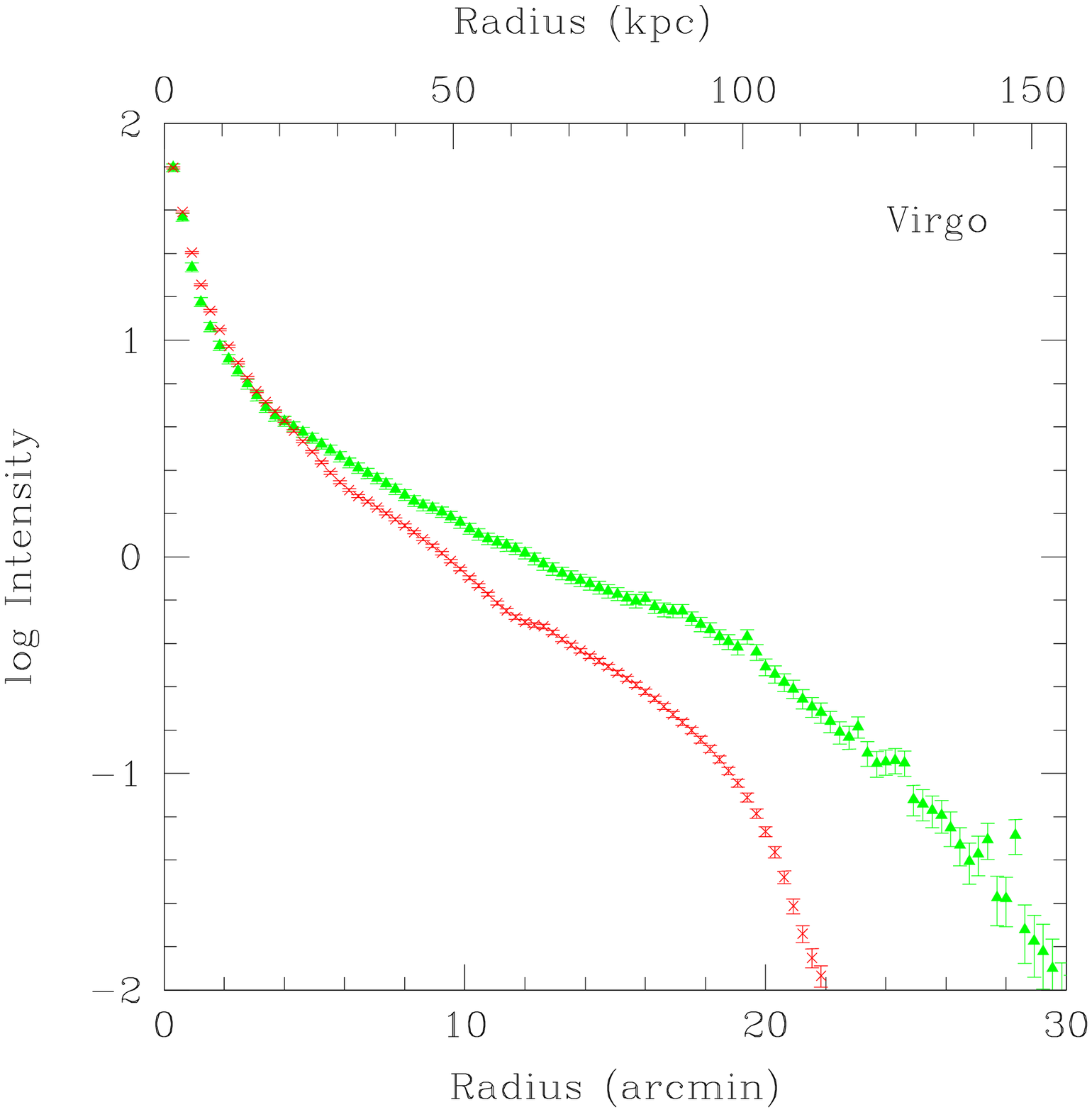,height=9truecm}
\caption[virgo_Ir_EUVX_fix_log.ps]{EUV and X-ray profiles
for Virgo. }
\label{fig:virgoprofil}
\end{center}
\end{figure}

\begin{figure}
\begin{center}
\psfig{figure=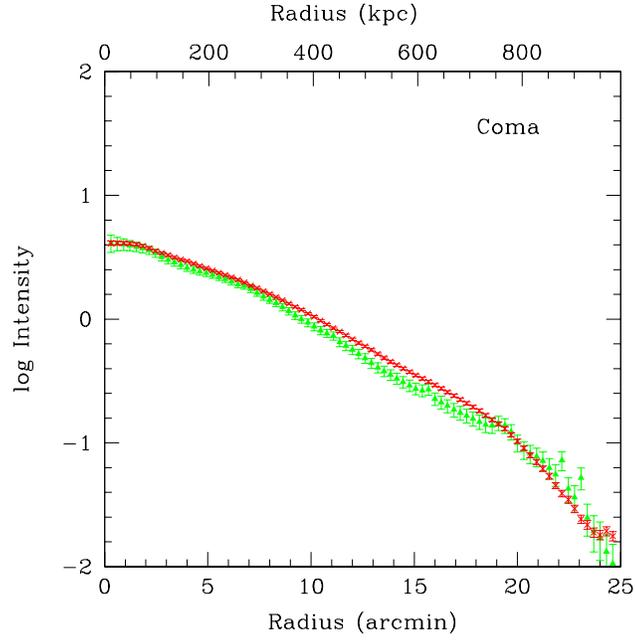,height=9truecm}
\caption[coma_Ir_EUVX_fix_log.ps]{EUV and X-ray profiles
for Coma. }
\label{fig:comaprofil}
\end{center}
\end{figure}

\begin{table}
\centering
\caption{Kolmogorov Smirnov test on EUV and X-ray profiles of Coma: 
probability P that the profiles differ for concentric rings of various sizes.}
\begin{tabular}{|lc|}
\hline
Radius (arcmin) & P \\
\hline		    
$0<R<2$   & 0.0122 \\
$2<R<17$  & ($1-4.8\ 10^{-11}$) \\
\hline	
$0<R<5$   & 0.1625 \\
$5<R<17$  & ($1-1\ 10^{-9}$) \\
\hline	
$0<R<10$  & 0.5809 \\
$10<R<17$ & 0.9999 \\
\hline
\end{tabular}
\label{tbl:kscoma}
\end{table}

The Coma cluster is the hottest cluster of our sample and behaves in a
very different way. The EUV and X-ray radii reach $\geq$17 and 20
arcmin respectively (Fig. \ref{fig:comaprofil}), somewhat larger than
reported e.g. by \citet{Bowyer3}, and corresponding to $\geq$660 and
780 kpc respectively.  The EUV and X-ray profiles appear to be very
similar, at least in the inner 10 arcminutes, as confirmed by the K-S
test; the probability that the two profiles differ varies with radius:
it is low in the very center and becomes large further out, as
illustrated in Table \ref{tbl:kscoma}.

\begin{figure}[h]
\begin{center}
\psfig{figure=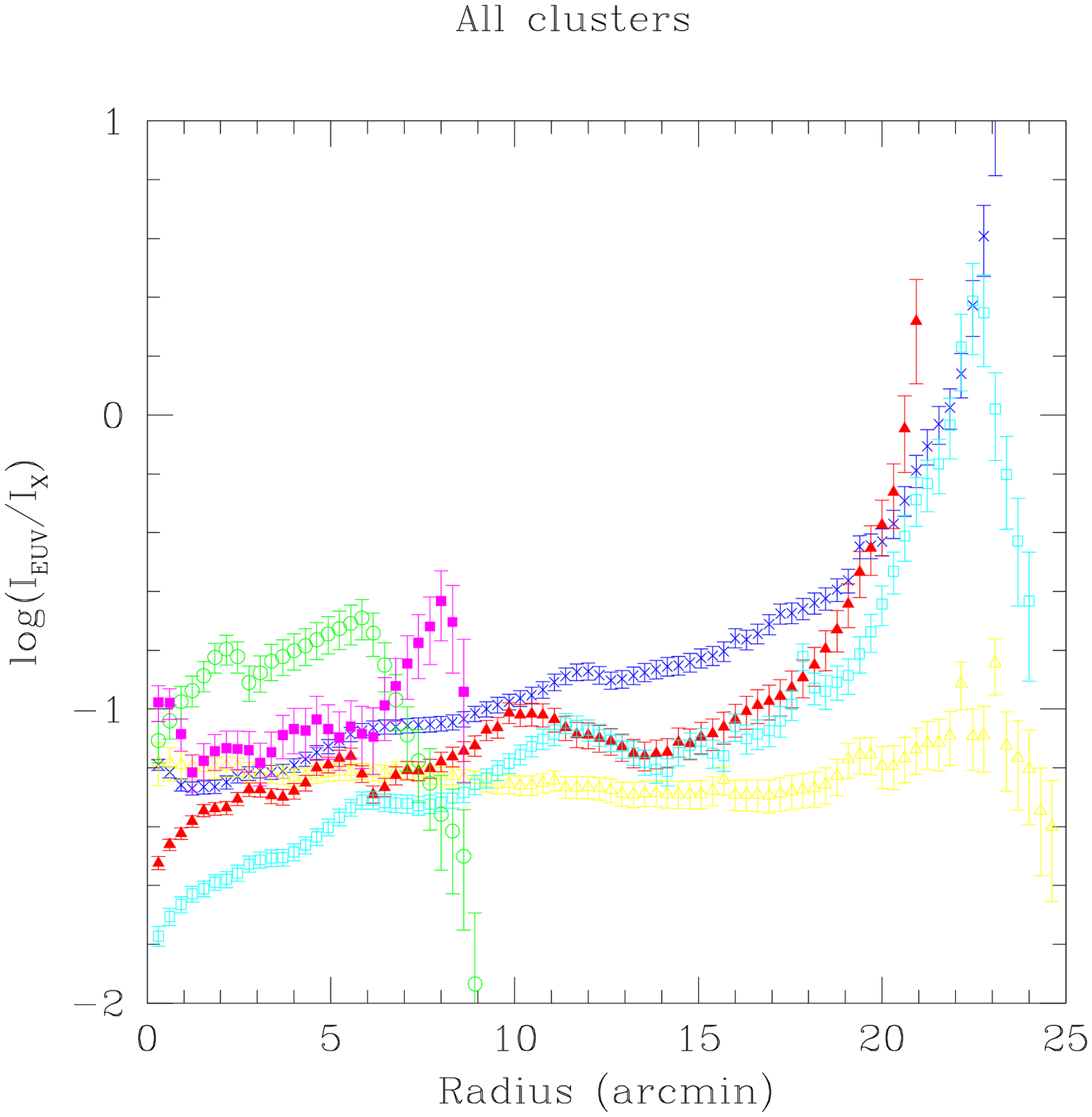,height=12truecm}
\caption[all_lograppEUVX_rmin.ps]{EUV and X-ray profiles for all six
clusters in our sample, with the radius expressed in arcminutes. 
The following symbols are used: 
A~1795: filled red triangles;
A~2199: empty light blue squares; 
A~4038: filled magenta squares; 
A~4059: empty green circles; 
Virgo: dark blue crosses; 
Coma: empty yellow triangles.
}
\label{fig:allrminprofil}
\end{center}
\end{figure}

\begin{figure}[h]
\begin{center}
\psfig{figure=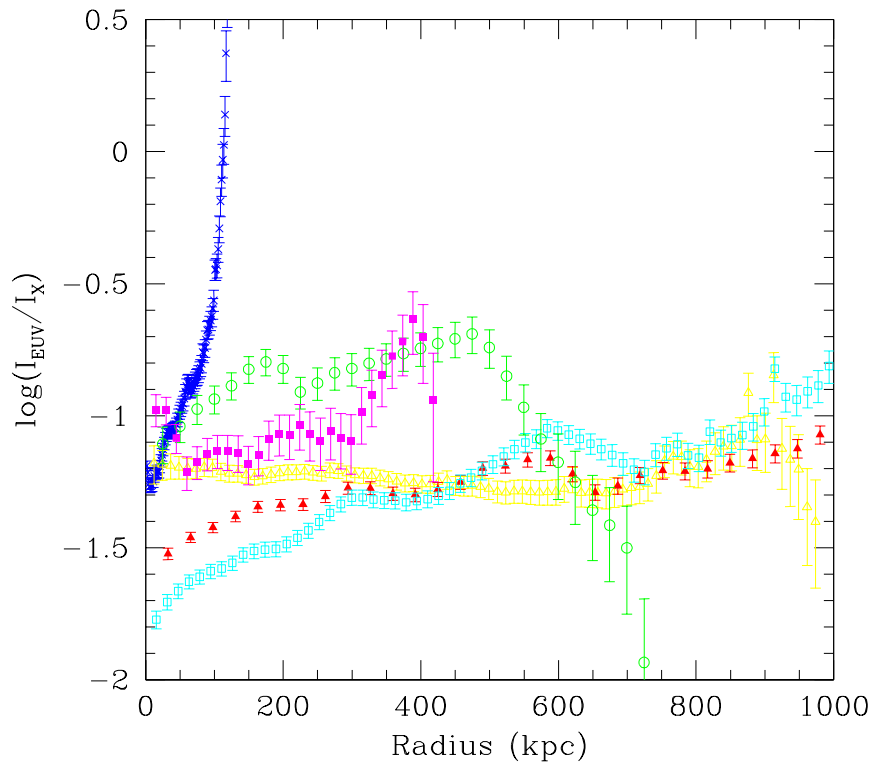,height=12truecm}
\caption[all_lograppEUVX_rkpc.ps]{EUV and X-ray profiles for all six
clusters in our sample, with the radius expressed in kpc. }
\label{fig:allrkpcprofil}
\end{center}
\end{figure}

In order to compare all six clusters, the ratios of the EUV to X-ray
intensities in concentric ellipses are shown in
Figs. \ref{fig:allrminprofil} and \ref{fig:allrkpcprofil} for radii
expressed in arcminutes and in kpc respectively.

As mentioned above, we can see that four clusters out of six appear to
extend both in the EUV and in X-rays at least up to 17 arcminutes,
that is notably further than previously reported for EUV images. The
two clusters which do not extend as far from the center (when the radii
are expressed in arcminutes) are A~4038 and A~4059. This small
physical extent is confirmed for A~4038 when the EUV to X-ray
ratio is plotted in physical units (kpc). However, the physical extent
of the EUV emission of A~4059 is 500-600 kpc, and therefore is not
small.

On the other hand, the radial extent of the EUV and X-ray emission in
Virgo is comparable to that of the other clusters when the radius is
expressed in arcminutes, but becomes extremely small in physical
units, due to the very small redshift of the cluster: slightly more
than 100 kpc, that is of the order of the size of the central galaxy M87
\citep{Bohringer}.

\section{Conclusions}

The fact that the EUV and X-ray emission profiles markedly differ in
five clusters out of the six of our sample clearly indicates that at
least in these five clusters the EUV emission cannot be only due to
the low energy tail of bremsstrahlung emission from the hot gas which
accounts for the X-ray emission.

A test will obviously be to see if the existence of a soft EUV excess
over bremsstrahlung emission from the hot X-ray emitting gas is
confirmed by XMM.

\bibliography{moriond}

\end{document}